\documentclass{article}
\usepackage{graphicx}
\usepackage{amsfonts}
\usepackage{amsmath}
\usepackage{epsfig}
\usepackage{natbib}
\usepackage[geometry]{ifsym}
\usepackage{xcolor}
\usepackage{float}
\usepackage{array, color}
\usepackage{amssymb}
\usepackage{lmodern}
\usepackage{bm}
\usepackage{url}
\title{Swimming of a ludion in a stratified sea}

\author{P. Le Gal$^\dag$, B. Castillo Morales$^{\ddag}$, S. Hernandez Zapata$^{\ddag}$,\\
 G. Ruiz Chavarria$^{\ddag}$\\
 $^\dag$ Aix-Marseille Universit\'e, CNRS, Centrale Marseille, IRPHE, 49 rue F. Joliot Curie,\\
13384 Marseille, C\'edex 13, France\\
$^{\ddag}$Departamento de F\`{\i}sica, Facultad de Ciencias,\\
 Universidad Nacional Autonoma de M\'{e}xico, 04510, M\'{e}xico}

\date{\today}

\begin{document}

\maketitle

\begin{abstract}
We describe and model experimental results on the dynamics of a "ludion" - a neutrally buoyant body - immersed in a
layer of stably stratified salt water. By oscillating a piston inside a cylinder
communicating with a narrow (in one of its horizontal dimensions) vessel containing the stably stratified layer
of salt water, it is easy to periodically vary the hydrostatic pressure of the fluid. The ludion or Cartesian diver, initially
positioned at its equilibrium height and free to move horizontally, can then oscillate vertically when forced by the pressure oscillations. Depending on the ratio of the forcing frequency to the Brunt-V\"{a}is\"{a}l\"{a} frequency
of the stratified fluid, the ludion can emit its own internal gravity waves that we measure
by a classical Particle Image Velocimetry technique. Our experimental results describe first the resonance of the vertical
motions of the ludion when excited at different frequencies. A theoretical oscillator model is
then derived taking into account added mass and added friction coefficients and its predictions are compared
to the experimental data. Then, for the larger oscillation amplitudes, we observe and describe a bifurcation towards free horizontal
motions. Although the internal gravity wave frequencies are affected by the Doppler shift induced by the horizontal displacement velocities, it
seems that, contrary to surface waves associated with Couder walkers \citep{Couder} they are not
the cause of the horizontal swimming. This does not however, exclude possible interactions between the ludion and internal gravity waves and possible hydrodynamic quantum analogies to be explored in the future.

\end{abstract}

\section{Introduction}
%\vspace{-2.5mm}

%\subsection{On the origin of the ludion}

\begin{figure}
\centering
(a)~~~~~\includegraphics[width=4cm]{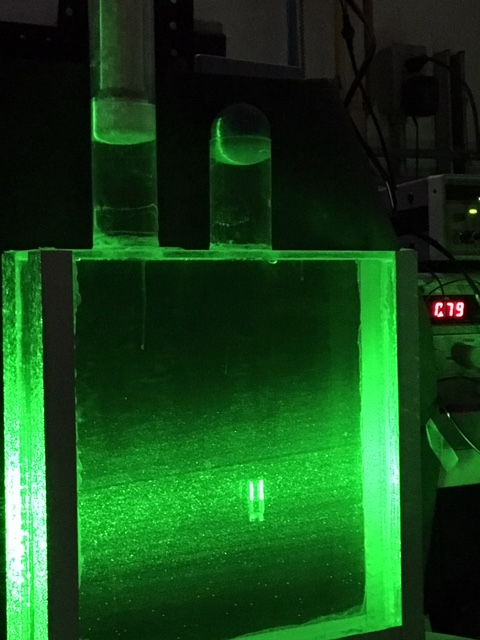}~~~~(b)\includegraphics[width=8cm]{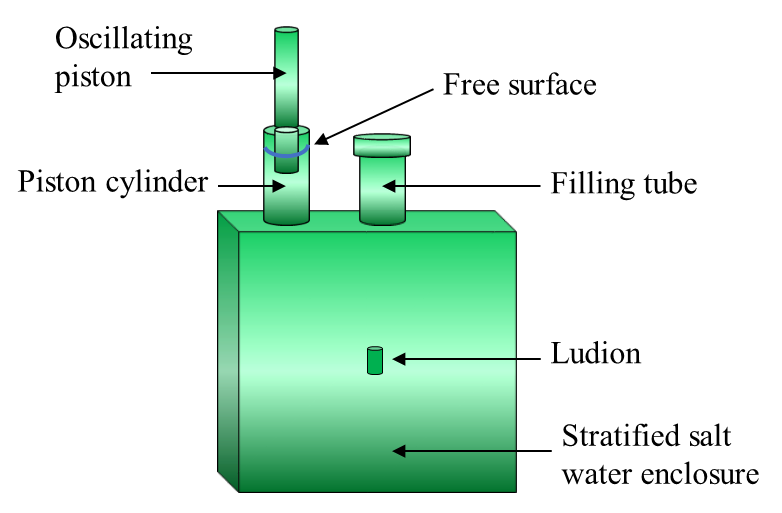}
\caption{a) Picture of the container with the stratified layer of salt water and the ludion at its stable height. b) Drawing of the container with its filling tube and piston cylinder. Because of the density stratification, the ludion floats at an
equilibrium position $z=0$ around which it is forced to oscillate vertically when the pressure is periodically changed by
moving the piston.}
\label{Fig1}
\end{figure}

The Cartesian diver (also called "ludion" in French) is a small object  denser than the water in which it is immersed but which encloses a
 pocket of air. By decreasing the pressure in the water, this air pocket expands, increasing the buoyancy force that opposes the weight
 of the diver that rises accordingly. In contrast, if the pressure  increases, the air compresses, causing the diver to sink. The first
 reference to such an object dates from 1648, when Raffaello Magiotti  published his work on the resistance of water to compression
 \citep{Magiotti}.  In this historical publication, we can find the very first drawings of a ludion credited to Magiotti.
 The notes that Magiotti had accumulated were destroyed during the great plague that raged in Rome in 1656 and of which he died.  Later, in
 the first half of the 18th century, John Theophilus Desaguliers, a  French-born philosopher, became a curator of the experiments of the
 Royal Society in London. Desaguliers wrote a book in experimental philosophy, in two volumes (1734 and 1744)
 in which he presented the  ludion \citep{Desaguliers}.  We will keep in the following both  appellations: the ludion and the diver.  However, it is not known how  the names "Cartesian diver" or "Cartesian devil" appeared and were  popularized (\cite{Ackerson} and references herein).  Unlike the
case of a homogeneous fluid, in which  the diver has an unstable equilibrium  position except in a small window of perturbation amplitudes \citep{Guemez}, if the diver is immersed in a stably stratified fluid it possesses a linearly stable equilibrium position as can be observed on figure \ref{Fig1}.

 If the pressure is varied  sinusoidally, the diver oscillates vertically around its stable position and behaves as an oscillator
 which can experience a resonance when the driving frequency is tuned. The first part of this study (section \ref{oscillo}) is
devoted  to the analysis of the ludion dynamics in the neighbourhood of this resonance. These calculations
 describe the main characteristics of the ludion dynamics including its added mass and friction coefficients. A series of experiments
 whose results are described in section \ref{reso} are devoted to the determination of the resonance which is compared to the
analytic prediction. If the forcing frequency is lower than the Brunt-V\"ais\"al\"a frequency $N$, the diver generates internal
 gravity waves that we characterize by Particle Image Velocimetry  (PIV) \citep{pivlab} and present in section \ref{IGWs}. At high
 amplitude response to the periodic forcing, i.e. when the forcing frequency is close to the resonant frequency, the ludion dynamics
 shows a bifurcation to free horizontal locomotion in a similar way as the vertically flapping wing of \cite{nvdb} or more recently
 as the oscillating spheroids simulated by \cite{Caulfield}. As soon as the oscillation amplitude is sufficiently large,
 this horizontal swimming appears regardless of the value of the forcing frequency, i.e. with or without the presence of internal gravity waves. The
 description of the bifurcation to this free dynamics is given in section \ref{swimm}. Finally, in the last section, some
 perspectives of this work possibilities for future research will be given. In particular, we mention the
 possible interaction of the ludion trajectories with its own internal gravity waves, which is reminiscent of the drops that bounce on the free
 surface of a vibrating liquid \citep{Couder,Perrard,Bush}.

\section{The forced damped oscillator model}
\label{oscillo}
%\vspace{-2.5mm}

In our experiments, a transparent acrylic rectangular container with dimensions 33~cm x 33~cm x 6~cm is filled with a linearly stratified layer of salt water. Its smallest dimension is in the horizontal plane.  Then a small hollow glass cylinder (the diver or ludion) (35~mm high for a diameter D= 12.5~mm) is introduced with care in order to preserve the density stratification. This diver (including its air pocket) was prepared to have a mean density intermediate between the minimum and
maximum density of the fluid and thus finds an equilibrium position in the stratified layer. The stratification is characterized by its Brunt-V\"ais\"al\"a frequency $N = \sqrt{- \frac{g}{\rho_0} \frac{d \rho}{dz}}$ where $z$ is the vertical coordinate, $g$ the gravity and $\rho$ the density of the fluid at level $z$. $\rho_0$ is
the density of the fluid at position $z=0$, chosen at the ludion equilibrium position. With the exception of an open pipe at its top wall (see figure \ref{Fig1}), the container is completely closed so that its pressure can be controlled by moving a piston inside this pipe which in consequence modifies the vertical level of the free surface of water inside the pipe and thus the pressure in the container. From the theoretical point of view, it is straightforward to write down the equation of motion of the ludion from the momentum conservation equation. In fact, this question has already been addressed in the past in the context of oceanography. \cite{Larsen} was the first to study the damped oscillations of a neutrally buoyant sphere in a stratified layer. He explicitly calculated the loss of power due to the radiation of internal gravity waves, neglecting the viscous friction. His results show that this radiative damping stops any small oscillations in few periods of oscillations. However, Larsen's calculation was criticized by \cite{Winant} who reconsidered the problem at the light of  experiments performed by \cite{Cairns} on the descent of neutrally buoyant floats in the ocean. \cite{Winant} considered for his analysis, a quadratic drag law as opposed to the linear law used by \cite{Larsen} that takes into account the gravity wave radiation. The conclusion drought by \cite{Winant} and observed in the experiments, was that at small displacements of the sphere, the damping term is due to internal gravity waves as predicted by \cite{Larsen} whereas one needs to incorporate in the equation of motion the quadratic drag term at large displacements as encountered in float descents from the sea surface.

In our experiments, the glass cylinder itself (of density $\rho_g$) occupies a volume $V_g$. Its mass is thus $M= V_g \rho_g$. The buoyancy force that opposes its weight is
$F_B = g \rho(z) (V_a + V_g)$, where $\rho(z)$ is the density of the fluid surrounding the ludion at position $z$ and $V_a$ the volume of the air bubble entrapped in the cylinder. Therefore, at equilibrium, supposed at $z=0$ and fluid density $\rho_0$, $M = \rho_0 (V_a + V_g)$. When changing the pressure in the container of stratified salt water by oscillating the piston up and down in its pipe, the initial volume of the air bubble $V_{a0}$ varies following a process that we suppose to be adiabatic:
$V_a(t,z) = V_{a0} \left( \frac{P_0}{P(t,z)}\right)^{1/\gamma}$
where $\gamma$ is the ratio of specific heats of air.
This hypothesis can be justified by the fact that the period of oscillation of the ludion will be around $3$ seconds, i.e. smaller than the the heat diffusion time in the air bubble of the order of $5$ seconds. Let us also remark that supposing an isothermal process will only
imply to take the value of $\gamma$ equal to $1$.
In the following, to obtain the equation of motion for the ludion that moves along the vertical axis of a distance $\xi$ versus the equilibrium position $z=0$, we will suppose that the change in volume of the air bubble trapped inside the diver affects only the buoyancy force. This is equivalent to the classical Boussinesq simplification of buoyant flows. The equilibrium pressure $P_0$ is modified by the addition of a small perturbation
$d p =\rho_0 ~ g ~ d h ~ cos(\omega t)$ created by moving up and down the piston in the
pipe, inducing respectively a decrease or an increase of the water level in the pipe of a quantity $d h$. We will see later that this pressure oscillation $d p$ will need to be supplemented by a surface tension contribution that will be added later when comparing the theory to the experimental data.
The motion of the ludion is then described by the following equation:

\begin{center}
\begin{equation}
M \frac{d^2 \xi}{dt^2} = - M g + F_B + F_A + F_H - \mu \frac{d \xi}{dt}
\end{equation}
\end{center}

$F_A$ is a hydrodynamical force that originates from the motion of water entrained by the displacement of the ludion. This force is classically written as:

\begin{center}
\begin{equation}
%F_A= - Real\left(C_A \rho_0 (V_{a0} + V_g)  \frac{d^2 \xi}{dt^2}\right)
F_A= - Real\left(C_A  M \frac{d^2 \xi}{dt^2}\right)
\end{equation}
\end{center}

\noindent with $C_A$ a complex added mass coefficient: $C_A = C_{Ar} + i C_{Ai}$. When the motion is periodic with an angular frequency $\omega$,
%$ \xi \sim e^{- i \omega t} $,
 $F_A$ can be split into two terms \citep{Lai,Ermanyuk,Voisin} that represent the added mass and the added friction
 to be incorporated in the equation of motion. Both coefficients $C_{Ar} $ and $C_{Ai}$ depend on the frequency $\omega$.

\begin{center}
\begin{equation}
%F_A= - Real(C_A) \rho_0 (V_{a0} + V_g)  \frac{d^2 \xi}{dt^2} + \omega Imag(C_A) \rho_0 (V_{a0} + V_g)  \frac{d \xi}{dt}
F_A= - C_{Ar} ~ M ~ \frac{d^2 \xi}{dt^2} + \omega ~C_{Ai} ~ M ~\frac{d \xi}{dt}
\label{Eq1}
\end{equation}
\end{center}
Therefore, the first term of $F_A$ will be added to the inertial term of equation \ref{Eq1} and the second term
will be a new dissipative term that complements the friction term $\mu \frac{d \xi}{dt}$ supposed to be for simplification of Stokes type because of the relative low value of the Reynolds number of the flows, as we will see later.
We should note however that
the study of the motions of bodies and of their drag in stratified layers has been the subject of an intense research (see the review by \cite{Magnaudet}) and it is today admitted that a supplementary drag (versus the drag exerted by the homogeneous fluid) originates in the buoyancy of the fluid entrained with the body \citep{Yick}.

$F_H$ is the history force, or the Basset force that appears when the motion is accelerated \cite{Basset,Boussinesq}. Very often, this term is omitted in the determination of the drag forces applied to moving bodies in particular when the fluid is stratified, the reason being that the motions are generally considered quasi-static \citep{Yick}. This memory force has been however explicitly calculated by \cite{Candelier} in the case of a sphere and in the limit of small Reynolds and Peclet numbers.
This term represents the diffusion of the vorticity in the boundary layers surrounding the body and is classically
written as:
\begin{center}
\begin{equation}
F_H= - Real\left(\eta ~ \int_{-\infty}^{0} \frac{d^2 \xi / d \tau^2}{\sqrt{t-\tau}} d\tau\right)
\label{Eq2}
\end{equation}
\end{center}
where $\eta$ is a complex coefficient $\eta = \eta_{r} + i \eta_{i}$.

In the same way we did for $F_A$, $F_H$ can be split into two terms thanks to
the periodicity of the motion that starts at $t=0$:
\begin{center}
\begin{equation}
F_H= - \eta_r ~ \int_{0}^{t} \frac{d^2 \xi / d \tau^2}{\sqrt{t-\tau}} d\tau~-~\omega~\eta_i ~ \int_{0}^{t} \frac{d \xi / d \tau}{\sqrt{t-\tau}} d\tau
\label{Fresnel}
\end{equation}
\end{center}

With an appropriate change of variables, the analytical expression of $F_H$ shows the appearance of the transcendental Fresnel integrals. However, if we study the behaviour of the ludion at large time, i.e. after tens of periods of oscillation, we can
consider the limits of the integrals as time goes to $\infty$ which are finite and known quantities and $F_H$ simplifies to:

\begin{center}
\begin{equation}
F_H= - \sqrt{\frac{\pi}{2\omega}} \left({\frac{\eta_r +\omega~\eta_i}{\omega^2}~\frac{d^2 \xi}{dt^2}+ \frac{\eta_r - \omega~\eta_i}{\omega}~\frac{d \xi}{
dt}}\right)
\label{Fresnel2}
\end{equation}
\end{center}

Therefore, as can be seen on equation \ref{Fresnel2}, this memory or history force $F_H$ can be incorporated in the
already existing added mass and added friction terms. This result was also used by \cite{Abad} in the case
of an oscillating sphere in an homogeneous fluid. In order not to overload the notation, we will keep the coefficients $C_{Ar}$ and $C_{Ai}$ knowing that they come from both the added mass force and the history force. Anyway, our experiments will not be able to desentangle the different origins of these terms.

To calculate $F_B$, we will expand its expression at first order in $\xi$ and $dp$ taking into account the variation of density
along the vertical axis and the change in volume due to the change in pressure:
\begin{center}
\begin{equation}
\rho (z)= \rho_0 +\left. \xi \frac{d \rho}{dz}\right|_0
\end{equation}
\end{center}

\begin{center}
\begin{equation}
V_a(z,t)= V_{a0} +  \left. \xi \frac{\partial V_a}{\partial z}\right|_0 + \left. dp \frac{\partial V_a}{\partial p}\right|_0
\end{equation}
\end{center}
 At first order, we obtain the expression of the buoyancy force:
 \begin{center}
\begin{equation}
F_B=g~ \rho_0 (V_{a0} + V_g)+ \xi \left( g~ (V_{a0}+V_g) \left.\frac{d \rho}{dz}\right|_0 + g~ \rho_0 \left. \frac{\partial V_a}{\partial z}\right|_0\right)+  g~ \rho_0 ~dp~ \left. \frac{\partial V_a}{\partial p}\right|_0
\end{equation}
\end{center}
The first term in the expression of $F_B$ will disappear as it is opposed to the weight of the ludion, and the equation of motion at first order reads:
\begin{center}
\begin{equation}
(1+ C_{Ar}) \frac{d^2 \xi}{dt^2} = \xi \left( \frac{g}{\rho_0}  \left.\frac{d \rho}{dz}\right|_0 + \frac{g}{V_{a0}+V_g}\left. \frac{\partial V_a}{\partial z}\right|_0 \right)+ \frac{g~dp~}{V_{a0}+V_g} \left. \frac{\partial V_a}{\partial p}\right|_0 - ( \frac{\mu}{M} + \omega ~C_{Ai}) ~\frac{d \xi}{dt}
\label{Eq2}
\end{equation}
\end{center}

The first term on the right hand side of equation \ref{Eq2} is simply equal to $-N^2 \xi$. The derivative of the volume in the second term can be rewritten as a function of the pressure, using the fact that the process is supposed to be adiabatic:
\begin{center}
\begin{equation}
\left. \frac{\partial V_a}{\partial z} \right|_0 = \frac{- V_{a0}}{\gamma P_0} \left. \frac{\partial P_a}{\partial z} \right|_0
\end{equation}
\end{center}
that finally leads to
\begin{center}
\begin{equation}
\frac{g}{V_{a0}+V_g} \left. \frac{\partial V_a}{\partial z} \right|_0 = \delta~ \frac{g^2 \rho_0}{\gamma P_{0}} =  \omega_0^2
\end{equation}
\end{center}
with $\delta = 1 - \frac{\rho_0}{\rho_g}$ and $\omega_0^2 =  \delta~\frac{g^2 \rho_0}{\gamma~ P_{0}}$, $P_{0}$ being the reference pressure of the air pocket entrapped inside the ludion when this one is at its equilibrium position.
The forcing term of equation \ref{Eq2} is the third term on the right hand side and is equal to $ - \delta~ \omega_0^2 ~dh ~cos(\omega t)$ if the free surface position is periodically changed by an amplitude $dh$ by the piston oscillations.

Finally, the equation of motion of the ludion along the vertical coordinate takes the form at first order of a damped forced harmonic oscillator :
\begin{center}
\begin{equation}
 (1+ C_{Ar}) \frac{d^2 \xi}{dt^2} = \xi ( - N^2 + \omega_0^2) - \omega_0^2 ~dh ~cos(\omega t) - ( \frac{\mu}{M} + \omega ~C_{Ai}) ~\frac{d \xi}{dt}
 \label{osc}
 \end{equation}
\end{center}
  The eigenfrequency is proportional to $\sqrt{ N^2 - \omega_0^2}$. We see that in a non stratified fluid, i.e. when $ N =0$, we recover the fact that the equilibrium position of the ludion in a pure fluid is unstable at first order of the expansion. In fact \nocite{Guemez} have shown that
  at second order, the non linear terms induce a saddle-node bifurcation in a limited domain of height and pressure perturbations
   (a fold catastrophe to take the terminology used by \cite{Guemez}). This effect will be ignored in the following as the density stratification (if large enough) of the fluid makes our system stable at first order. It is then traditional to rename the damping coefficient by $ 2~ \lambda$ :
  \begin{center}
\begin{equation}
2~ \lambda = \frac{\frac{\mu}{M} + \omega ~C_{Ai}}{1+ C_{Ar}}
\end{equation}
\end{center}

\section{Experimental observation of the resonance}
\label{reso}

\begin{figure}
\begin{center}
\includegraphics[width=15 cm]{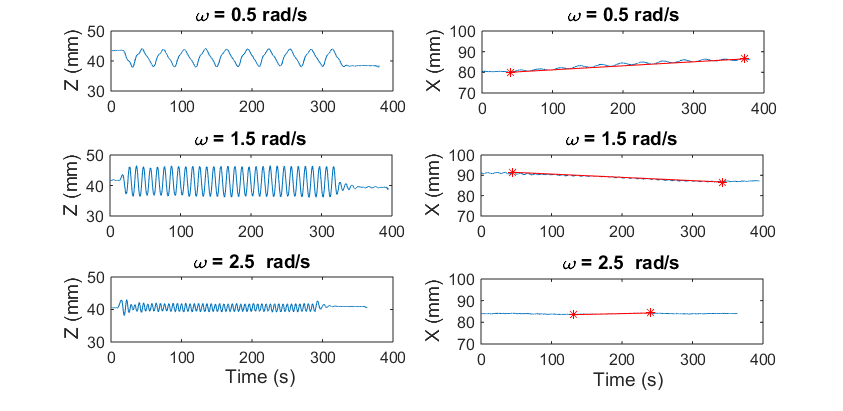}
\caption{Tracking of the ludion by video image analysis for three forcing frequencies with a Brunt-V\"ais\"al\"a frequency $N = 1.6~ rad/s$. left: vertical oscillations vs. time. right: horizontal excursion vs. time. red stars are the minimum and the maximum horizontal coordinates of the ludion. The red solid line between the stars is used to calculate the mean horizontal velocity between the stars.}
\label{Fig2}
\end{center}
\end{figure}
All of the following experiments were realized in the Physics Department of UNAM in Mexico. Before filling the container with stratified salt water, some micron size PIV particles are added to the fluid. The enclosure is then slowly filled using the double bucket technique \citep{Oster1965} to create the density stratification using salt (NaCl). A $5~cm$ thick homogeneous dense layer of salt (fresh) water is kept at the bottom (respectively at the top) of the container. As already explained, the ludion is then carefully immersed in the stratified fluid layer that is illuminated with a green laser sheet in order to record by a video camera (60 frames per second) the ludion oscillations as well as the particle motions in its neighborhood. In all of the experiments, the piston is translated sinusoidally in its cylinder at a chosen frequency by a precise stepper motor ($25000$ steps per rotation), provoking a rise or a descent of the water surface inside the piston cylinder. This change in pressure is a priori measured by the amplitude of the free surface oscillations which is in all the cases presented here equal to $1.4~cm$.
The forcing frequency is varied from $0.3~rad/s$ to $2.5~rad/s$ and each run is recorded for several minutes. The position of the ludion is then calculated by a specially designed tracking software based on the brightness of the image as the ludion reflects the laser light more intensively that the fluid background.

\begin{figure}
\begin{center}
\includegraphics[width=10 cm]{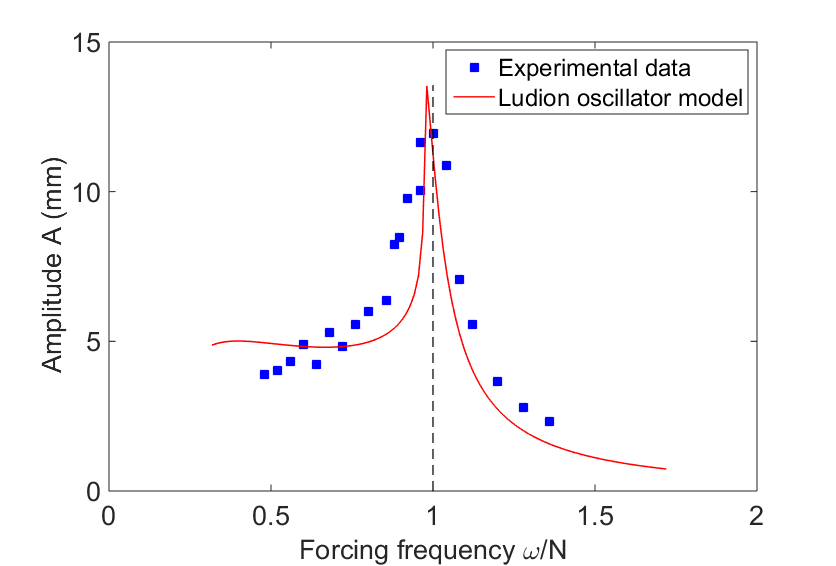}
\caption{Resonant curve of the ludion with a Brunt-V\"ais\"al\"a frequency $N = 1.6~ rad/s$. The data points (squares) are the vertical oscillation amplitudes $A$ collected from the experimental trajectories. The red solid line is the result of the analytical expression of the resonance of the oscillator that takes into account the added mass and added friction coefficients interpolated from the transient measurements whereas the forcing free surface elevation $dh$ is chosen equal to $2.8~ cm$ and the effective glass density as defined in the text, measured to be equal to $1445~kg/m^3$. There are no free coefficients in the theoretical model.}
\label{Fig3}
\end{center}
\end{figure}

Figure \ref{Fig2} presents three examples of the dynamics of the ludion. We will be first interested by the amplitude of the
vertical motions along the $Z$ axis which are illustrated in the left column of the figure. The horizontal motions (along the X axis) will be studied later in section \ref{swimm}. As can be observed, the amplitude $A$ of the vertical excursions is a function of the forcing frequency as expected by the resonant response of a damped forced oscillator. Figure \ref{Fig3} shows this behaviour. We observe also that the maximum amplitude is reached for a frequency slightly smaller than $N$ as expected from the model.

\begin{figure}
\begin{center}
a)~~\includegraphics[width=10 cm]{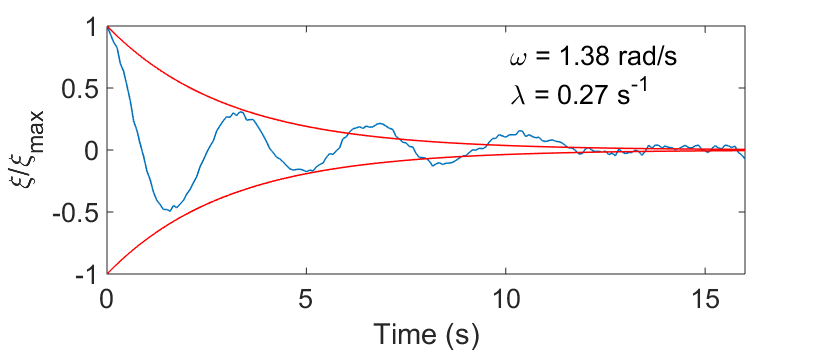}
b)~~\includegraphics[width=10 cm]{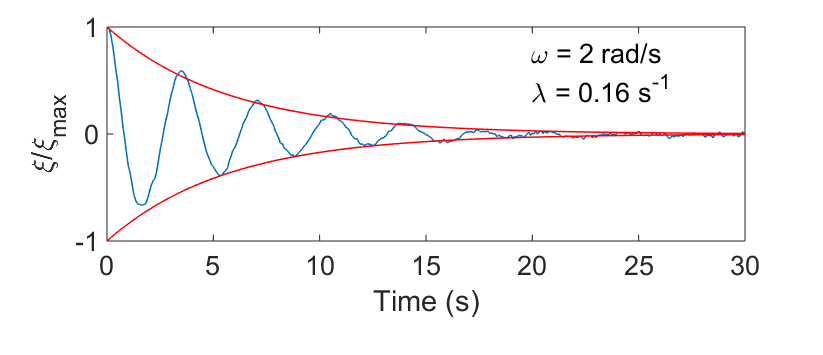}
\caption{Transient vertical oscillations of the ludion when the forcing is stopped ($N = 1.6~rad/s$). A linear fit of the logarithm of $\xi/\xi_{max}$, leads to the measurement of the damping coefficient. a)  $\lambda = 0.27~s^{-1}$ when frequency $\omega = 1.38~rad/s$. b) $\lambda = 0.16~s^{-1}$ when frequency $\omega=2~rad/s$.}
\label{Fig4}
\end{center}
\end{figure}

In order to get a comparison between our ludion oscillator model and the experimental data, we need to compute the different coefficients of equation \ref{osc}. First, the mass of the ludion is determined by measuring its weight, but it appears that this
mass needs to be completed by the mass of the water entrapped under the air pocket and transported with the glass cylinder. From the picture of the ludion we can guess that air and water are approximatively of equal volume inside the diver. This will lead to an effective glass density (mean density of the non compressible part constituted by glass and entrapped water in the cylinder) of $1445~kg/m^3$ i.e. smaller than the real glass density. The second term to be estimated is the real pressure changes experienced by the ludion. This was first estimated by the rise and fall of the water level in the piston pipe. But as already mentioned earlier, the changes in the water level need to be completed by a surface tension term. As there is a small gap of the order of $1~mm$ between the piston and the cylinder, this surface tension pressure is of the order of magnitude of the hydrostatic pressure. Thus, we calculate that the total change in pressure is in fact twice the $140~Pa$ estimated first. Therefore we will use a value of $2.8~cm$ that will lead to a good comparison with the experimental data. Then we can measure the added mass and friction coefficients by the study of the ludion damped oscillations after the forcing is stopped. This technique was already used by \cite{Ermanyuk2000} and we will use directly the analytical relations derived in their work. The first step to evaluate $C_{Ar}$ and $C_{Ai}$, is to determine the damping coefficient $\lambda$ from a best fit of the temporal evolutions. Figure \ref{Fig4} shows two examples of the damped oscillations of the ludion when the forcing is stopped. Note that contrary to the power-law decaying oscillations observed by \cite{Biro}, an exponential damping law fits properly our experimental data at least on the rather limited number of periods of oscillation that we measured.

\begin{figure}
\begin{center}
\includegraphics[width=10 cm]{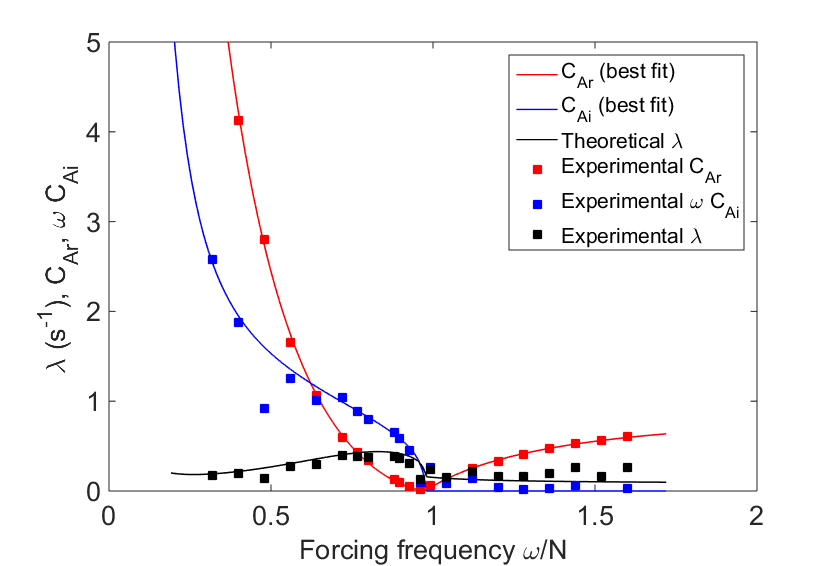}
\caption{Evolution of the damping mass and friction coefficients with the reduced forcing frequency $\omega /N$. First the typical time scale of the damping is extracted from the exponential fit of the transients (black squares). Then using the formula explicitly derived by \cite{Ermanyuk2000}, we can calculate the real part $C_{Ar}$ and the imaginary part $\omega~C_{Ai}$ of the complex added mass. The solid curves are then heuristic best fits calculated from modified analytical formula given by \cite{Voisin} for an oscillating sphere (see Appendix).}
\label{Fig5}
\end{center}
\end{figure}

We performed this analysis for the whole range of forcing frequencies. Figure \ref{Fig5} shows the evolution of $\lambda$ as a function of $\omega$. In particular, we recover the typical shape of the damping coefficient evolution with its modification by the emission of gravity waves as it is calculated theoretically (see \cite{Lai,Ermanyuk,Voisin}). We clearly see that when $\omega$ is larger than $N$, the damping is only due to the viscous effects with a typical frequency estimated around $\frac{\mu}{M} = 0.16 s^{-1}$ in order to recover $\omega~C_{Ai} = 0$ when $\omega$ is larger than $N$. From this experimental data, and using the formulas given by \cite{Ermanyuk2000}, we can explicitly write the added mass and friction coefficients:

\begin{center}
\begin{equation}
C_{Ar}= \left|\left|{\frac{N^2 - \omega_0^2}{\omega^2+\lambda^2} - 1}\right|\right|
\end{equation}
\end{center}
\begin{center}
\begin{equation}
\omega~C_{Ai} = 2~\lambda~\frac{N^2 - \omega_0^2}{\omega^2+\lambda^2} - \frac{\mu}{M}
\end{equation}
\end{center}

The experimental values of $C_{Ar}$ and $\omega~C_{Ai}$ are also plotted on figure \ref{Fig5}. It is remarkable that these experimental data points possess the same trends that the theoretical calculations of \cite{Lai,Ermanyuk,Voisin} for spheres and horizontal cylinders. We were able to fit them by the modified analytical expressions obtained by \cite{Voisin} for a vertically oscillating sphere (See Appendix). These fitting curves are
then used to recalculate the friction coefficient $\lambda$ to be used in the oscillator model in order to reproduce the resonant response of the ludion.
%The fact that our measures of $\lambda$ are close to the theoretically obtained curve is a proof of the consistency.
As can be observed in figure \ref{Fig3}, the analytical prediction of the model (red line) gives without any free coefficients, a satisfactory prediction of the experimental resonance curve. In particular the slight deviation of the resonant frequency versus the Brunt-V\"ais\"al\"a frequency $N$ is visible. Note also the presence of a "shoulder" on the left wing of the resonant curve which is reminiscent of the increase of the power loss by the radiation of internal gravity waves.

\section{A gravity wave generator}
\label{IGWs}
\begin{figure}
%\centering
~\hspace{-.5cm}a)\includegraphics[width=6.7 cm]{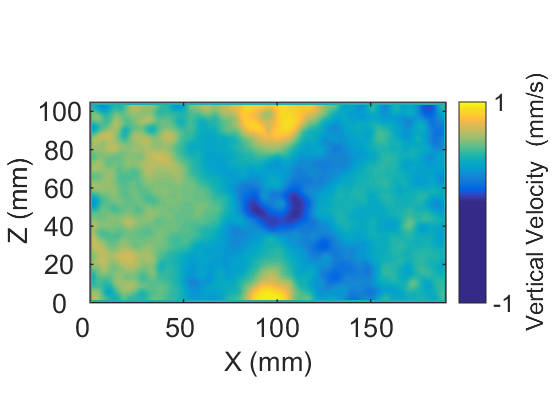}~~b)\includegraphics[width=6.7 cm]{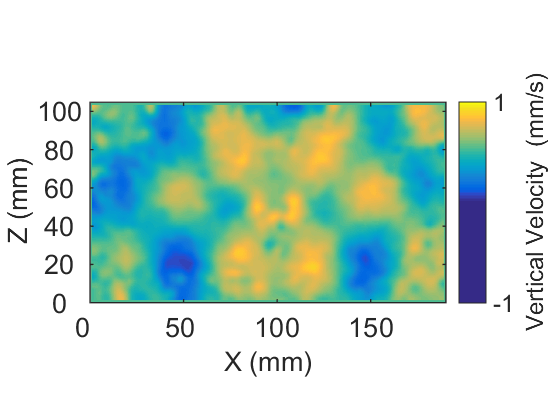}
\caption{ a) The Saint Andrew's cross of the internal gravity waves emitted by the ludion measured in a vertical plane by PIV (forcing frequency $\omega = 1.38~rad/s$ and and $N = 1.6 rad/s$) 30 periods after the beginning of the oscillation (Vertical velocity field). b) The chessboard pattern after multi reflections have occurred, about 10 periods of oscillation later.}
\label{Fig6}
\end{figure}
%\vspace{-2.5mm}
When the forcing frequency is less than the Brunt-V\"ais\"al\"a frequency, internal waves accompany the diver in its oscillations. Seeding the salt water with micron sized particles, it is possible to measure the fluid motions around the ludion by PIV \cite{pivlab}.
 Figure \ref{Fig6}-a) shows in color the vertical velocity field in a vertical plane. A typical Saint Andrew's cross is clearly visible for a forcing frequency $\omega$ equal to 1.38 rad/s some 30 periods after the oscillations started. However, as
 shown on Figure \ref{Fig6}-b), multi-reflections of the gravity waves on the lateral walls and on the
 density gradients at the bottom and top homogeneous layer frontiers make the wave pattern to look like a chessboard pattern
 reminiscent of an underlying eigenmode. In the present case, the eigenfrequency of mode ($n_x = 4, n_y = 1, n_z = 2$) where $n_j$ is the number of wavelengths in direction $j$,  is also equal to $ 1.38~ rad/s$.  Note however that this correspondence is obviously frequency dependant and does not occur in general. Moreover, we  did not find at this point of our experimental work any correlation between the amplitude of the vertical oscillations as described in section \ref{reso} and the possible excitation of a global eigenmode in the container.

\begin{figure}
\centering
~\hspace{-.5cm}\includegraphics[width=10 cm]{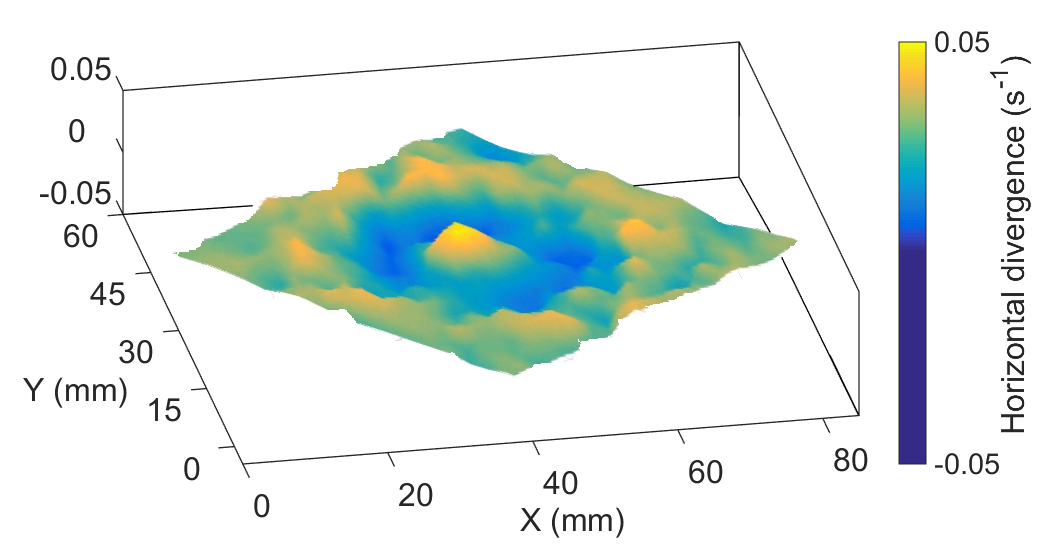}
\caption{The horizontal divergence of the velocity field in a horizontal plane 1 cm above the ludion (forcing frequency $\omega =~1.26 rad/s$ and $N = 1.6 rad/s$).
}
\label{Fig7}
\end{figure}

  As it is well known, the arms of the Saint Andrew's cross are wave conical iso-phase surfaces. These cones can be observed in horizontal planes as presented by the horizontal divergence of the velocity field in Figure \ref{Fig7} measured $1~cm$  above the highest position of the ludion.

\section{The diver can also swim}
\label{swimm}

%\vspace{-2.5mm}

As visible on the right column plots of Figure \ref{Fig2}, in some of the experiments we have observed that the ludion moves or swims along the permitted horizontal axis X. We recall that in the other direction, the ludion is  confined between two vertical walls distant of $6~cm$.
To describe this motions, we have detected the maximum displacements of the ludion along the horizontal axis during the recording time of the run. Then we calculated the corresponding mean velocity represented by the slope of the red lines in Figure \ref{Fig2}. We have analysed two sets of experiments with different density stratifications leading respectively to $N= 1.6~rad/s$ and  $N=~2.3 rad/s$. As already noted in the introduction, this kind of horizontal locomotion associated  with vertically oscillating body has already been observed in particular for the vertically flapping wing of \cite{nvdb} or more recently for the oscillating spheroids simulated by \cite{Caulfield} where it is demonstrated that this horizontal propulsion is directly linked to a symmetry breaking
 in the vortical flow pattern generated at each oscillation. These last studies follow in fact the seminal experimental work of \cite{Tatsuno} on the flow generated by horizontal oscillations of a cylinder in a homogeneous fluid. \nocite{Tatsuno} produced a classification of the different flows they observed as function of two non dimension parameters : the Keulegan-Carpenter number $KC = 2\pi A/D$ and the Stokes number $\beta = \omega D^2 / 2\pi \nu $ where $\nu$ is the kinematic viscosity of water. In particular, these authors determined a transition between symmetric and asymmetric flows (called Regime D in their article). It is this critical threshold that \nocite{Caulfield} as well as other authors refer to as the
transition towards propulsion. Therefore, and even if the ludion is a small vertical cylinder (and not a sphere) and even if  all of these previously cited experiments and calculations were realized in non stratified fluids, we will also refer to the \nocite{Tatsuno} curve in the following. Note moreover that in the case of stratified fluids, the oscillating body is always attached in a way or another to an oscillating device. This is the case for instance for the experiments of \nocite{Lin} where a classification diagram is also presented in the ($KC$, $\beta$) plane for a Froude number larger than $0.20$ and compared to the results of \nocite{Tatsuno}. In particular, a critical threshold is found above which internal gravity waves are emitted by the periodic oscillations of a sphere. The generation of internal gravity waves by oscillating bodies in stratified fluids (see for instance \cite{Bruce}) is of course a long standing history starting from the description of the "herring bone" pattern by \cite{Mowbray} and \cite{Stevenson}. More recently,\cite{Chashechkin} have also documented the flow patterns (cumulative jets and wave beams) around oscillating spheres and their singular features using a Schlieren technique.

\begin{figure}
\centering
\includegraphics[width=12 cm]{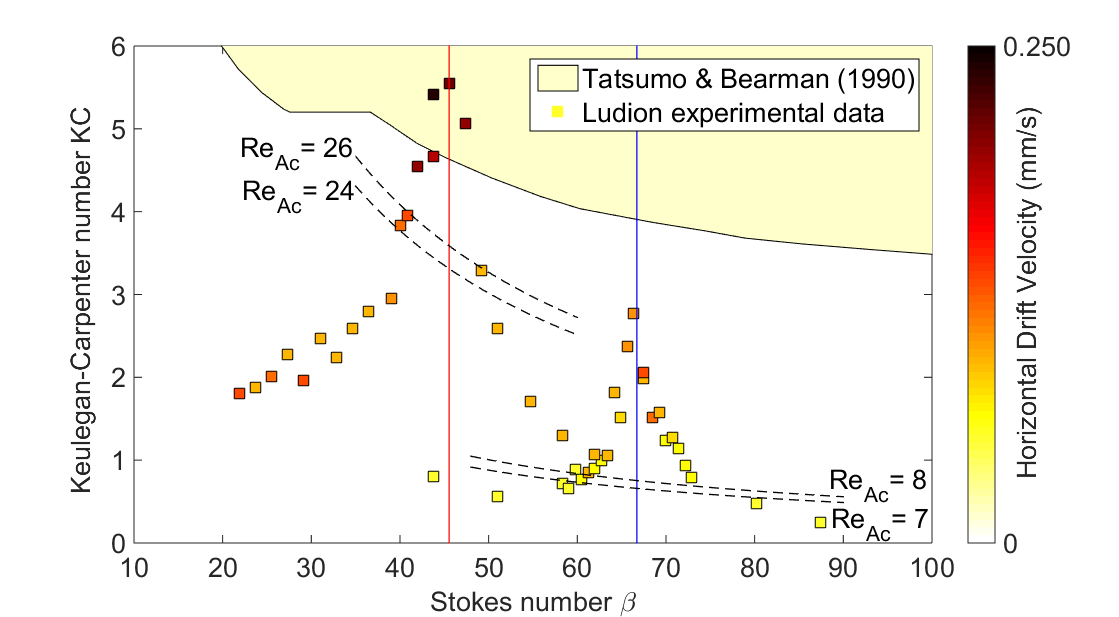}
\caption{The resonant curves in the ($\beta, KC$) plane for both $N=1.6~rad/s$ (red vertical line) and $N=2.3~rad/s$ (blue vertical line). The colors of
the data points are functions of the horizontal velocity as given in the colorbar. Dashed lines refer to instability thresholds given by constant
values of flapping Reynolds numbers determined on the bifurcation diagrams of figure \ref{Fig9}.}
\label{Fig8}
\end{figure}

Using the non dimension parameters $KC$ and $\beta$ as defined above, we can plot in figure \ref{Fig8} the resonant curves for both Brunt-V\"{a}is\"{a}l\"{a} frequencies $N= 1.6~rad/s$ and  $N= 2.3~rad/s$. In both cases, the Froude number defined as $Fr= \omega A / N D$ stays between $0.05$ and $1$. On this figure, the data points are colored following the intensity of the horizontal velocity experienced by the ludion. As can be observed, the more intense vertical displacements lead to a larger horizontal velocity. We have also represented by dotted lines the estimated thresholds for the apparition of horizontal excursions. By defining a flapping Reynolds number $Re_A = \beta~KC/2~\pi$ we can describe the transition between oscillations with and without horizontal motions. We observe on figure \ref{Fig8} that these thresholds are lower than the one detected by \cite{Tatsuno}. Therefore we can conclude that density stratified flows associated to vertical oscillations of a body seem to be more sensitive to flow symmetry breaking and in consequence to horizontal propulsion.

\begin{figure}
\centering
\includegraphics[width=12 cm]{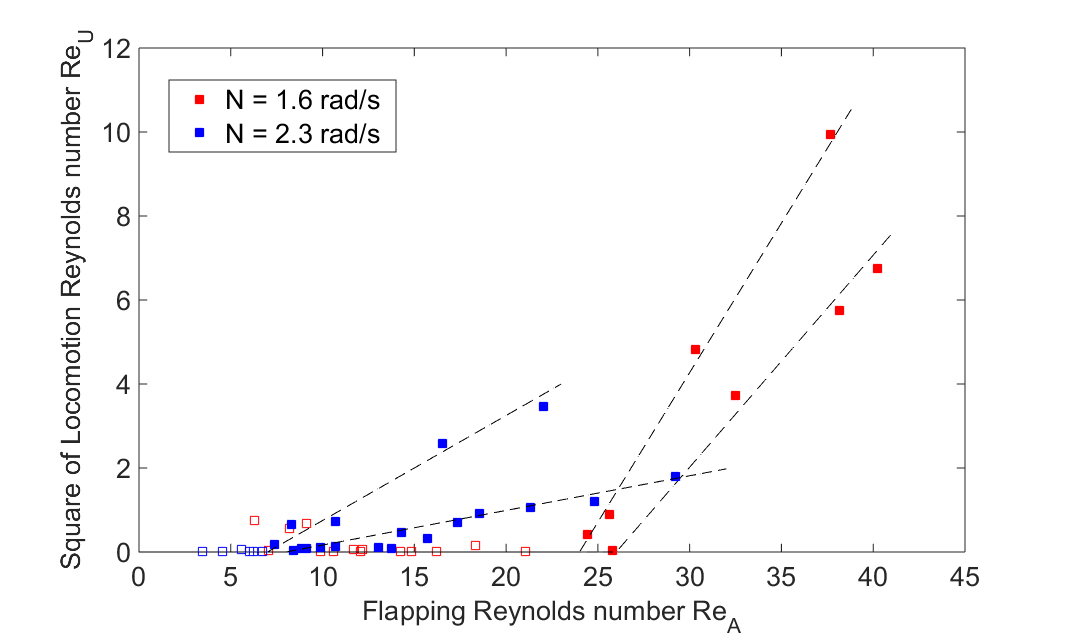}
\caption{Pitchfork bifurcations towards horizontal motions of the ludion. We observe the different branches that bifurcate at different threshold values. Here the horizontal velocities have been converted into locomotion Reynolds numbers. For each experiment, two bifurcated branches are observed (solid symbols).}
\label{Fig9}
\end{figure}

 To compare with the numerical observations of \cite{Caulfield}, we have calculated for each experiment the locomotion Reynolds number $Re_U = V_H D /\nu$. As described in figure \ref{Fig9}, in both cases we observe pitchfork supercritical bifurcations with the amplitude
of the horizontal velocity $V_H$ (proportional to the locomotion Reynolds number) varying as the square root of the distance to thresholds. A particular feature of our study is the observation for both sets of experiments of two bifurcated branches having slightly different threshold values. These branches seem not to be correlated with the left or right wing of the resonant curves and might correspond to different flow symmetries as described in \cite{PRFDeng}. Note also that if the critical flapping Reynolds number that we observe have similar values that those for the oscillating spheroids of \cite{Caulfield}, the locomotion Reynolds number is several order of magnitudes lower in our case, revealing the poor efficiency of our swimmer - a feature that was obviously not aimed in our study.

\begin{figure}
\centering
\includegraphics[width=12 cm]{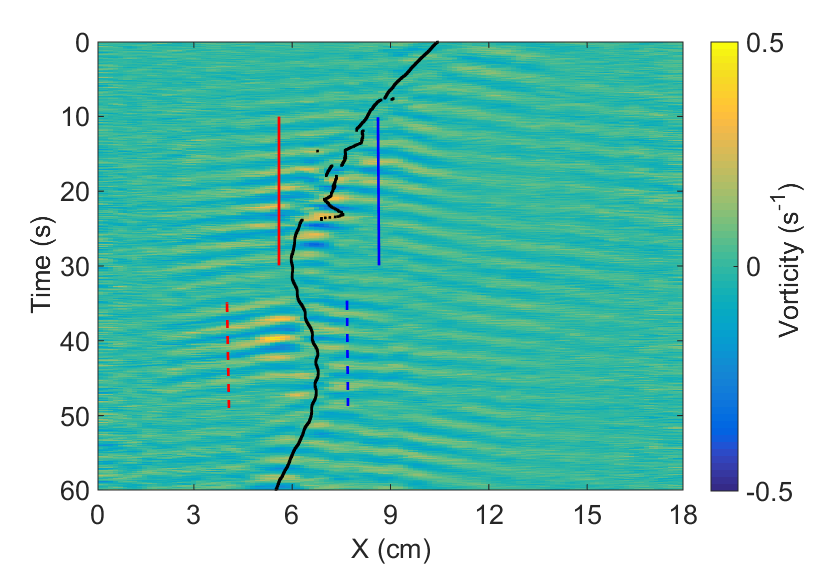}
\caption{Space-time diagram of the PIV vorticity field together with the ludion trajectory as calculated by our tracking software (black line). Four lines are drawn ahead or behind the ludion, along which the Fourier spectra of the vorticity fields are calculated and presented on figure \ref{Fig11}.
 }
\label{Fig10}
\end{figure}

To characterize the interaction between the ludion and the internal gravity waves it emits, we calculate the space-time diagram along an horizontal line located just above the diver using the PIV vorticity fields in the plane $(X, Z)$. Figure \ref{Fig10} presents this diagram where the propagation of the waves are visible on each side of the trajectory which is as expected the source of the waves. On the figure, we have also reported in black the trajectory of the ludion as determined by the tracking study presented in section \ref{reso}. Except from limited local defects which are due to spurious laser reflections on the glass cylinder, the correspondence between the detected trajectory and the trace of the ludion on the PIV map is good.

As it is expected from a wave emitter moving at a given velocity, the frequency of the gravity waves should be shifted by a Doppler shift given by the product $k_x~V_H$ where $k_x$ is the wavenumber of the gravity waves in the $X$ direction and $V_H$ the horizontal instantaneous locomotion velocity of the ludion. On the same figure, we have also drawn four lines on each side of the trajectory. Between the solid red and blue lines the ludion swims towards the left whereas between the dash red and blue line it swims towards the right. Therefore each of the red (respectively blue) lines are ahead (respectively behind) of the ludion. The velocity $V_H$ between the solid lines is equal to approximately to $0.8~mm/s$. Note that this instantaneous velocity is larger
 than the mean velocity (equal to $0.23~mm/s$) calculated on a longer period of time and reported of figure \ref{Fig8}. The wavelength of the gravity
 waves can also be estimated on figure \ref{Fig10} around $40~mm$ . This value corresponds to a wavenumber equal to $k_x=0.157~rad/mm$. These values give then a Doppler shift equal to $0.12~rad/s$.

\begin{figure}
\centering
\includegraphics[width=12 cm]{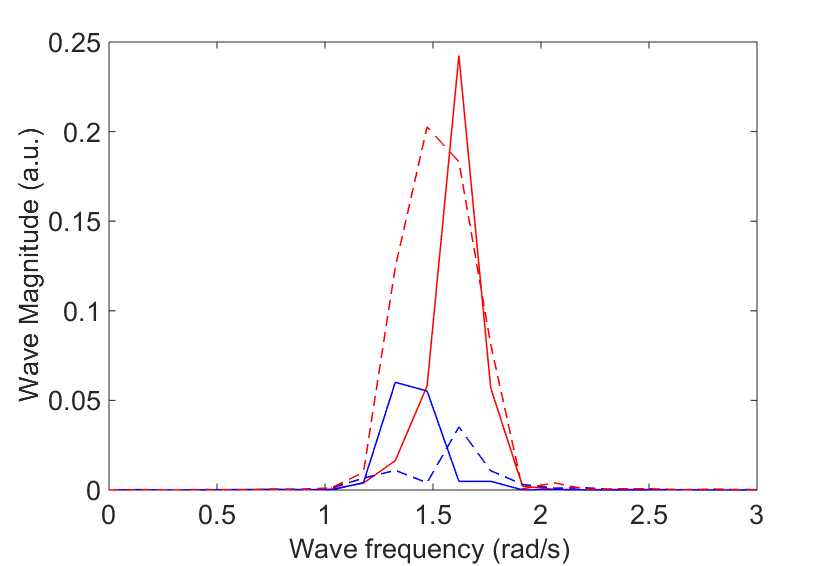}
\caption{Fourier spectra of the vorticity field along the lines drawn on figure \ref{Fig10}. As can be observed, the angular frequency of the gravity waves ahead of the ludion is higher that of the waves behind as expected by a Doppler shift. In this example, $\omega = 1.50~rad/s$ and $ N = 1.60~rad/s$.}
\label{Fig11}
\end{figure}

To check this prediction, we calculate the Fourier spectra of the vorticity along the four lines of figure \ref{Fig10}.  They are represented on figure \ref{Fig11} keeping for each spectrum the color and line styles identical to those of figure \ref{Fig10}. As can be observed, the angular frequency of the waves emitted ahead of the trajectory (and crossing the red solid line and the blue dash line)  are slightly larger than the frequency of the waves emitted behind (and crossing the red dash line and the blue solid line). As can be checked on figure \ref{Fig11}, the separation between the maxima of the spectra is equal to $0.23~rad/s$, leading to a shift of $\pm0.12~rad/s$ between the wave angular frequencies and the forcing frequency $\omega$ validating as expected the presence of a Doppler shift. Note moreover that associated to this frequency shift there is also a slight increase of the wave amplitude behind the ludion. We have at this stage of the study no interpretation of this effect.

\section{Conclusions and perspectives}
%\vspace{-2.5mm}

This work is the first step of the study of the swimming of a vertically oscillating but free -in the horizontal plane-  neutrally buoyant body (the ludion) in a density stratified fluid. All of the presented experiments were performed in a rectangular but narrow container in order to constraint the horizontal displacements in a single direction. We have been able to measure and model the resonance of the ludion taking into account the added mass and added friction terms in the equation of motion. We have also shown that the Basset (or memory) term can be incorporated in the added coefficients because the motion is harmonic and supposed to last for a sufficiently long time. In particular, we have measured the damping rate of the ludion oscillations during transients. As expected, the radiation of internal gravity waves increases the energy loss and thus the damping term. This emission of these gravity waves for forcing frequencies less than the Brunt-V\"{a}is\"{a}l\"{a} frequency has been characterized by PIV measurements.
Then a bifurcation towards locomotion (the swimming) is observed above a critical flapping Reynolds number measured at values lower
than what was expected from previous works performed in homogeneous fluids. Finally we showed that this swimming of the ludion is accompanied by a Doppler shift of the wave frequency.

\begin{figure}
\centering
\includegraphics[width=10 cm]{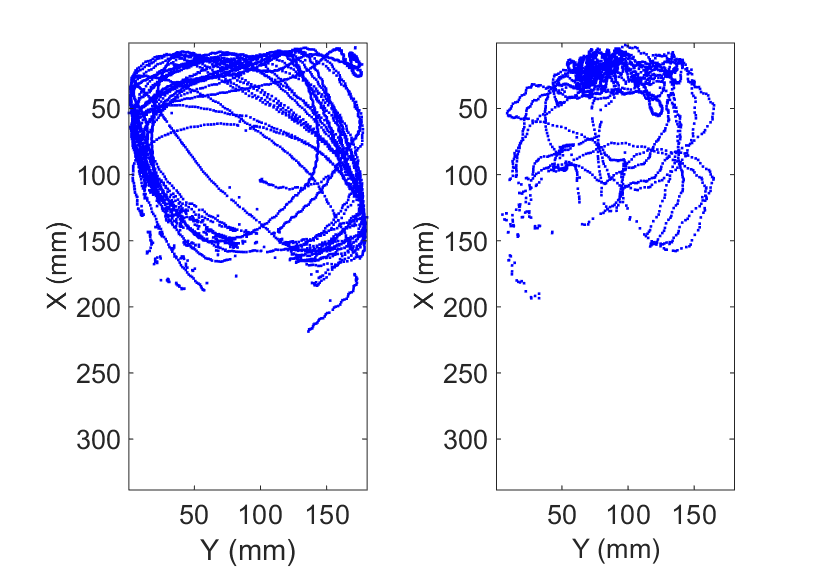}
\caption{Two examples of trajectories of the ludion in a larger container for $ N = 1.29~rad/s$. On the left, nearly 2 hours recording with $\omega = 1.23~rad/s$ and on the right, 3/4 hour recording at  $\omega = 1.20~rad/s$.}
\label{Fig12}
\end{figure}

Our experiments on the swimming of the ludion in a stratified fluid are of course reminiscent of the drops that bounce on the free surface of a vibrating liquid \citep{Couder,Bush}. Some differences between the bouncing drops and our system should however be pointed. First, the ludion experiences in our case a forced resonance and not a parametric instability as it is the case for the drops bouncing on a subcritical wave pattern of the Faraday instability.  Moreover, in our system, the waves are continuously emitted and in interaction with the body which is not the case for the drops as they leave the fluid surface. However, it appears that in our case the threshold to locomotion is not caused by the waves as we observed horizontal displacements even when the forcing frequency is higher than the Brunt-V\"{a}is\"{a}l\"{a} frequency. This is different from the walkers which start to move because of the presence of their waves. Another important difference with Couder walkers, is the propagative character of the internal gravity waves emitted by the ludion whereas the surface waves induced by the bouncing drops are stationary.  Even if not the cause of swimming, the gravity waves should interact with the ludion through the surface integral of the pressure they exert on it. One of the result of the waves effect on the ludion motions is the presence of some added mass and added friction terms evidenced in the equation of motion and that we characterized through our experimental measurements. As it is the case for the bouncing drops, history or memory terms exist but we were not able to evaluate them as we can not desentangle them from the added mass and friction terms.  We have moreover highlighted the presence of a Doppler shift on the gravity wave frequency as they are emitted by a moving source. This will cause a phase shifts between the waves emitted by the front and by the rear sides of the diver, inducing
pressure differentials and horizontal forces that at their turn should modify the ludion motions. The feed back loop, between the swimming of the ludion and the generated wave pattern it generates is certainly quite subtle.

 \begin{figure}
\centering
a) \includegraphics[width=5.0 cm, height = 5.5 cm]{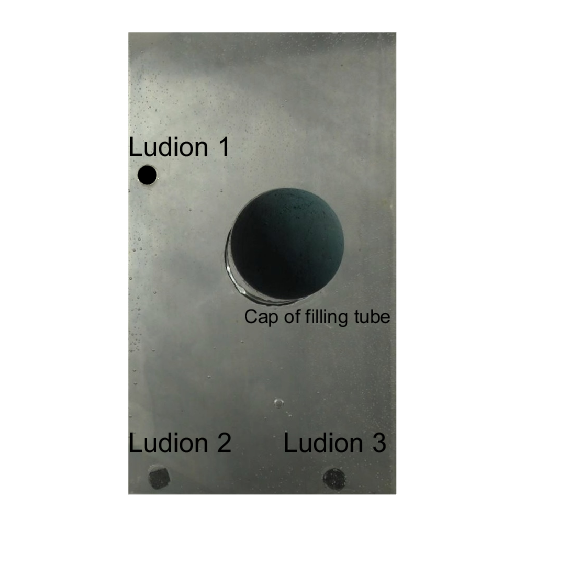}~b) \includegraphics[width=5.6 cm,height = 5.7 cm]{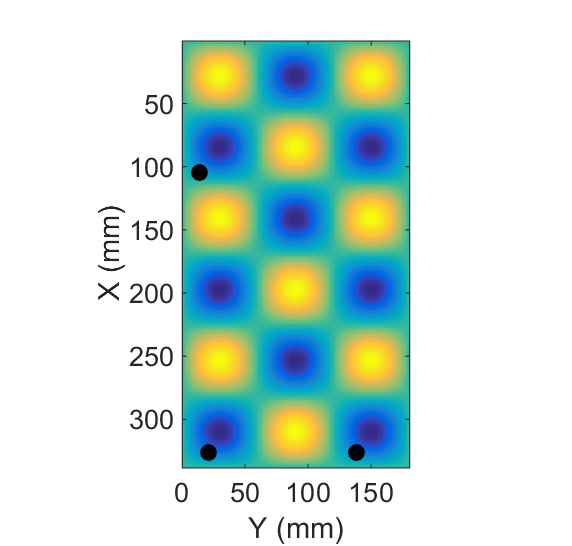}
\caption{An example where three ludions were introduced in the experimental set-up. a) A top view of the container where after a transient, the three ludions were motionless in the horizontal plane. These oscillate vertically in phase as they are forced with the pressure oscillation with  $\omega = 2.136~rad/s$ and $ N = 2.19~rad/s$. b) superimposed on the 3 ludions positions, the eigenmode at a frequency equal to  $2.135~rad/s$ with half a wavelength in the vertical direction, three in the $Y $direction and six in the $X$ direction. }
\label{Fig13}
\end{figure}

To open perspectives on these works, we present in figure \ref{Fig12} two trajectories of the ludion in a horizontal plane of a larger container and
using the same experimental procedure as that described above for the experiments in the narrow container.
In these two examples we could record from the top, long temporal series of swimming. In both cases, the ludion moves along trajectories confined
on one half of the container. In the first case, the trajectories wind up along elliptic loops whereas in the second case the ludion describes loops
close to trefoils \citep{Perrard} before slowing down, attracted to a motionless position.  In fact, it happens quite often during these first experiments in an extended geometry, that the ludion was attracted by some fixed locations and stayed there motionless. These locations might correspond to suitable spatial positions inside an underlying eigenmode, proving in a sense that the internal gravity waves can indeed interact with the ludion and might drive some of its motions. This is the case for instance in the experiments presented on figure \ref{Fig13}-a) where after the introduction of three divers in the stratified layer, they were all of them attracted to three places that as can be observed on figure \ref{Fig13}-b), seem to correspond to trapped locations of an underlying eigenmode that possesses an eigenfrequency very close to the forcing frequency. Note that an important constraint to respect is the identical phase between each ludion and the mode itself as the three divers are locked in phase to the forced pressure oscillations.
Therefore, and even if at this stage of the study it is not yet possible to precisely quantify the effect of the internal gravity waves on the motions and trajectories of the ludion, our observations encourage us to continue the search for possible hydrodynamics analogies with undulatory mechanics, in the quest for a new fluid mechanics example of a wave-particle duality as already beautifully observed and explored by \cite{Couder}, \cite{Perrard}  and \cite{Bush} for the Couder walkers.

\section{Appendix}
Figure \ref{Fig5} presents the evolution with the forcing frequency of the experimental values of $C_{Ar}$ and $\omega~C_{Ai}$ that are calculated
from the experimental values of $\omega$ and $\lambda$. As can be checked on the theoretical calculations of \cite{Lai},\cite{Ermanyuk}, and \cite{Voisin}, the general trends of these variations are largely recovered. In these articles, it is shown in particular that $C_{Ar}$ and $\omega~C_{Ai}$ depend on the aspect ratio of the considered oscillating body. As the ludion is a finite size cylinder that oscillates vertically along its axis - a case never considered theoretically - we will modify the analytical expressions derived for the sphere by \cite{Voisin} in order to fit our experimental data points. These will then be used to calculate the "theoretical $\lambda$" of figure \ref{Fig5} necessary to compute the resonant curve.
The complex added mass coefficient $C_z$ for a sphere oscillating vertically at frequency $\omega$ in a stratified fluid of Brunt-V\"{a}is\"{a}l\"{a} frequency $N$ is \citep{Voisin}:
\begin{center}
\begin{equation}
C_z(\omega /N)=\left(1-\frac{N^2}{\omega^2}\right) \frac{B(\omega /N)}{1-B(\omega /N)}
\end{equation}
\end{center}
 with
 \begin{center}
\begin{equation}
B(\omega /N)=(\omega^2 /N^2) \left[1-\left(1-\frac{N^2}{\omega^2}\right)^{1/2} \arcsin(N/\omega)\right]
\end{equation}
\end{center}

We will then defined heuristically the real and imaginary part of $C_A$ by:\\

$\bullet$~if $\omega \leq N$,  $C_{Ar} = 10~Real(C_z)^{1.4}$~and ~$C_{Ai} = 1.44~Imag(C_z)^{0.6}+ 0.22~(\omega/N)^{-2.4}- 0.16 $,\\

$\bullet$~if $\omega \geq N$,  $C_{Ar} = 3.8~Real(C_z)^{2.1}$ ~and~ $C_{Ai} = Imag(C_z)$,\\

 where the different coefficients have been determined by best fit trial and error method. These analytical expressions are then used
 in the ludion oscillator model.

\section*{Acknowledgments}

We thank B. Favier, G. Verhille, B. Voisin and S. Perrard for fruitful discussions during the course of this study. This research was financed by DGAPA-UNAM, under contract
PAPIIT IN-114218 "Vorticidad y ondas (internas y de superficie) en din\'amica de fluidos"  and IRPHE of Aix Marseille University.

 \section*{Declaration of interests}
The authors report no conflict of interest.

\bibliographystyle{jfm}
\bibliography{biblioludion}

\begin{thebibliography}{32}
\expandafter\ifx\csname natexlab\endcsname\relax\def\natexlab#1{#1}\fi

\bibitem[Abad \& Souhar(2004)]{Abad}
{\sc Abad, M. \& Souhar, M.} 2004 Effects of the history force on a
  noscillating rigid sphere at low reynolds number. {\em Exp. Fluids\/} {\bf
  36}, 775--782.

\bibitem[Ackerson(2020)]{Ackerson}
{\sc Ackerson, Bruce~J.} 2020 Cartesian diver plus. {\em The Physics Teacher\/}
  {\bf 58}~(2), 84--85.

\bibitem[Basset(1888)]{Basset}
{\sc Basset, A.B.} 1888 {\em Treatise of Hydrodynamics\/}, , vol.~2. Deighton
  Bell and Co, Cambridge.

\bibitem[Bir{\'{o}} {\em et~al.\/}(2008)Bir{\'{o}}, G\'{a}bor~Szab{\'{o}},
  Gy\"{u}re, J\'{a}nosi \& T\'{e}l]{Biro}
{\sc Bir{\'{o}}, I., G\'{a}bor~Szab{\'{o}}, K., Gy\"{u}re, BB., J\'{a}nosi,
  I.~M. \& T\'{e}l, T.} 2008 Power-law decaying oscillations of neutrally
  buoyant spheres in continuously stratified fluid. {\em Physics of Fluids\/}
  {\bf 20}~(5), 051705.

\bibitem[Boussinesq(1885)]{Boussinesq}
{\sc Boussinesq, J.} 1885 Sur la r\'esistance qu'oppose un fluide ind\'efini au
  repos sans pesanteur au mouvement vari\'e d'une sph\`ere solide qu'il mouille
  sur toute sa surface quand les vitesses restent bien continues et assez
  faibles pour que leurs carr\'es et produits soient n\'egligeables. {\em C. R.
  Acad. Sci. Paris\/} {\bf 100}, 935--937.

\bibitem[Bush(2015)]{Bush}
{\sc Bush, J.W.M.} 2015 Pilot-wave hydrodynamics. {\em Ann. Rev. Fluid Mech.\/}
  {\bf 47}, 269--292.

\bibitem[Cairns {\em et~al.\/}(1979)Cairns, Munk \& Winant]{Cairns}
{\sc Cairns, J., Munk, W. \& Winant, C.} 1979 On the dynamics of neutrally
  buoyant capsules; an experimental drop in lake tahoe. {\em Deep-Sea Res. A\/}
  {\bf 26}~(4), 369–--381.

\bibitem[Candelier {\em et~al.\/}(2014)Candelier, Mehaddi \&
  Vauquelin]{Candelier}
{\sc Candelier, F., Mehaddi, R. \& Vauquelin, O.} 2014 The history force on a
  small particle in a linearly stratified fluid. {\em Journal of Fluid
  Mechanics\/} {\bf 749}, 184--200.

\bibitem[Chashechkin \& Prikhod'ko(2007)]{Chashechkin}
{\sc Chashechkin, Yu.~D. \& Prikhod'ko, Yu.~V.} 2007 Regular and singular flow
  components for stimulated and free oscillations of a sphere in continuously
  stratified liquid. {\em Doklady Physics\/} {\bf 52}~(5), 261--265.

\bibitem[Couder {\em et~al.\/}(2005)Couder, Proti\`{e}re, Fort \&
  Boudaoud]{Couder}
{\sc Couder, Y., Proti\`{e}re, S., Fort, E. \& Boudaoud, A.} 2005 Walking and
  orbiting droplets. {\em Nature\/} {\bf 437}, 238--.

\bibitem[Deng \& Caulfield(2018)]{Caulfield}
{\sc Deng, J. \& Caulfield, C.~P.} 2018 Horizontal locomotion of a vertically
  flapping oblate spheroid. {\em J. Fluid Mech.\/} {\bf 840}, 688--708.

\bibitem[Deng {\em et~al.\/}(2017)Deng, Xue, Mao \& Caulfield]{PRFDeng}
{\sc Deng, J., Xue, J., Mao, X. \& Caulfield, C.~P.} 2017 Coherent structures
  in interacting vortex rings. {\em Phys. Rev. Fluids\/} {\bf 2}, 022701.

\bibitem[Desaguliers(1744)]{Desaguliers}
{\sc Desaguliers, J.T.} 1744 {\em A Course of Experimental Philosophy\/}. {\em
  A Course of Experimental Philosophy\/} vol.~2. W. Innys.

\bibitem[Ermanyuk(2000)]{Ermanyuk2000}
{\sc Ermanyuk, E.V.} 2000 The use of impulse response functions for evaluation
  of added mass and damping coefficient of a circular cylinder oscillating in
  linearly stratified fluid. {\em {Exp. Fluids}\/} {\bf 28}, 152--159.

\bibitem[Ermanyuk \& Gavrilov(2003)]{Ermanyuk}
{\sc Ermanyuk, E.V. \& Gavrilov, N.} 2003 Force on a body in a continuously
  stratified fluid. part 2. sphere. {\em {J. Fluid Mech.}\/} {\bf 494}, 33--50.

\bibitem[Flynn {\em et~al.\/}(2003)Flynn, Onu \& Sutherland]{Bruce}
{\sc Flynn, M.~R., Onu, K. \& Sutherland, B.~R.} 2003 Internal wave excitation
  by a vertically oscillating sphere. {\em Journal of Fluid Mechanics\/} {\bf
  494}, 65--93.

\bibitem[G\"{u}\'{e}mez {\em et~al.\/}(2002)G\"{u}\'{e}mez, Fiolhais \&
  Fiolhais]{Guemez}
{\sc G\"{u}\'{e}mez, J., Fiolhais, C. \& Fiolhais, M.} 2002 The cartesian diver
  and the fold catastrophe. {\em American Journal of Physics\/} {\bf 70}~(7),
  710--714.

\bibitem[Lai \& Lee(1981)]{Lai}
{\sc Lai, R.Y.S. \& Lee, C-M.} 1981 Added mass of a spheroid oscillating in a
  linearly stratified fluid. {\em Int. J; Engng Sci.\/} {\bf 19}, 1411--1420.

\bibitem[Larsen(1969)]{Larsen}
{\sc Larsen, L.~H.} 1969 Oscillations of a neutrally buoyant sphere in a
  stratified fluid. {\em Deep-Sea Res.\/} {\bf 16}~(6), 587--603.

\bibitem[Lin {\em et~al.\/}(1994)Lin, Boyer \& Fernando]{Lin}
{\sc Lin, Q., Boyer, D.~L. \& Fernando, H. J.~S.} 1994 Flows generated by the
  periodic horizontal oscillations of a sphere in a linearly stratified fluid.
  {\em Journal of Fluid Mechanics\/} {\bf 263}, 245--270.

\bibitem[Magiotti(1648)]{Magiotti}
{\sc Magiotti, R.} 1648 {\em Renitenza certissima dell'acqua alla
  compressione\/}. Moneta.

\bibitem[Magnaudet \& Mercier(2020)]{Magnaudet}
{\sc Magnaudet, J. \& Mercier, M.~J.} 2020 Particles, drops, and bubbles moving
  across sharp interfaces and stratified layers. {\em Annual Review of Fluid
  Mechanics\/} {\bf 52}~(1), 61--91.

\bibitem[Mowbray \& Rarity(1967)]{Mowbray}
{\sc Mowbray, D.~E. \& Rarity, B. S.~H.} 1967 The internal wave pattern
  produced by a sphere moving vertically in a density stratified liquid. {\em
  Journal of Fluid Mechanics\/} {\bf 30}~(3), 489--495.

\bibitem[Oster(1965)]{Oster1965}
{\sc Oster, G.} 1965 Density gradients. {\em Scientific American\/} {\bf
  213}~(2), 70--79.

\bibitem[Perrard {\em et~al.\/}(2014)Perrard, Labousse, Miskin, Fort \&
  Couder]{Perrard}
{\sc Perrard, S., Labousse, M., Miskin, M., Fort, E. \& Couder, Y.} 2014
  Self-organization into quantized eigenstates of a classical wave-driven
  particle. {\em Nature Communications\/} {\bf 5}, 3219.

\bibitem[Stevenson(1969)]{Stevenson}
{\sc Stevenson, T.~N.} 1969 Axisymmetric internal waves generated by a
  travelling oscillating body. {\em Journal of Fluid Mechanics\/} {\bf 35}~(2),
  219--224.

\bibitem[Tatsuno \& P.W.(1990)]{Tatsuno}
{\sc Tatsuno, M. \& P.W., Bearman} 1990 A visual study of the flow around an
  oscillating circular cylinder at low keulegan–carpenter numbers and low
  stokes numbers. {\em Journal of Fluid Mechanics\/} {\bf 211}, 157--182.

\bibitem[Thielicke \& Stamhuis(2014)]{pivlab}
{\sc Thielicke, W. \& Stamhuis, E.} 2014 {PIV}lab--towards user-friendly,
  affordable and accurate digital particle image velocimetry in matlab. {\em
  Journal of Open Research Software\/} {\bf 2}~(1).

\bibitem[Vandenberghe {\em et~al.\/}(2004)Vandenberghe, Zhang \&
  Childress]{nvdb}
{\sc Vandenberghe, N., Zhang, J. \& Childress, S.} 2004 Symmetry breaking leads
  to forward flapping flight. {\em Journal of Fluid Mechanics\/} {\bf 506},
  147--155.

\bibitem[Voisin(2007)]{Voisin}
{\sc Voisin, B.} 2007 {Added mass effects on internal wave generation}. In {\em
  {5th International Symposium on Environmental Hydraulics}\/}, p. Compact disk
  of the Conference. Tempe, United States.

\bibitem[Winant(1974)]{Winant}
{\sc Winant, C.~D.} 1974 The descent of neutrally buoyant floats. {\em Deep-Sea
  Res.\/} {\bf 21}~(6), 445--453.

\bibitem[Yick {\em et~al.\/}(2009)Yick, Torres, Peacock \& Stocker]{Yick}
{\sc Yick, K.~Y., Torres, C.~R., Peacock, T. \& Stocker, R.} 2009 Enhanced drag
  of a sphere settling in a stratified fluid at small reynolds numbers. {\em
  Journal of Fluid Mechanics\/} {\bf 632}, 49--68.

\end{thebibliography}

\end{document}